\pdfoutput=1
\documentclass[sigplan,screen,authorversion,nonacm]{acmart} 
\usepackage[utf8]{inputenc}
\usepackage{amsmath,amsfonts,amssymb}
\usepackage{mathtools}
\usepackage{color}
\usepackage{tikz}
\usepackage{hyperref}
\usepackage{phdalgo}
\usepackage[capitalize]{cleveref}
\usepackage{numprint}
\usepackage{accents}
\usepackage{paralist}
\usepackage{adjustbox}
\usepackage{booktabs}
\newcolumntype{P}[1]{>{\centering\arraybackslash}p{#1}}
\newcolumntype{M}[1]{>{\centering\arraybackslash}m{#1}}
\usepackage{siunitx}
\usepackage{microtype}

\newtheorem{conjecture}{Conjecture}
\newtheorem{theorem}{Theorem}
\newtheorem{proposition}[theorem]{Proposition}
\newtheorem{lemma}[theorem]{Lemma}

\newcommand{\etc}{etc}
\newcommand{\ie}{i.e.}
\newcommand{\eg}{e.g.}
\newcommand{\cf}{cf.}

\newcommand{\eps}{\varepsilon}
\newcommand{\R}{\mathbb{R}}

\newcommand{\Pcal}{\mathcal{P}}

\newcommand{\Ptilde}{\widetilde{\mathcal{P}}}
\newcommand{\btilde}{\widetilde{b}}

\newcommand{\Coq}{\textsc{Coq}}
\newcommand{\CoqEAL}{\textsc{CoqEAL}}

\newcommand{\geqlex}{\mathrel{\geq_\mathrm{lex}}}

\newcommand{\leqlex}{\mathrel{\leq_\mathrm{lex}}}

\newcommand{\CoqPolyhedra}{\textrm{Coq-Polyhedra}}

\newcommand{\MathCompShort}{MathComp}
\newcommand{\scalar}[2]{\langle #1, #2 \rangle}

\newcommand{\trans}[1]{#1^T}

\newcommand{\Minimize}{\textrm{Minimize}}

\newcommand{\SubjectTo}{\textrm{subject to}}

\newcommand{\floor}[1]{\left\lfloor #1 \right\rfloor}
\newcommand{\ceil}[1]{\left\lceil #1 \right\rceil}

\newcommand{\Gvert}{G_\mathrm{vert}}
\newcommand{\Gbases}{G_\mathrm{bases}}
\newcommand{\Glex}{G_\mathrm{lex}}
\newcommand{\Id}{\mathrm{Id}}

\definecolor{dkblue}{rgb}{0,0.1,0.5}
\definecolor{lightblue}{rgb}{0,0.5,0.5}
\definecolor{dkgreen}{rgb}{0,0.4,0}
\definecolor{dk2green}{rgb}{0.4,0,0}
\definecolor{dkviolet}{HTML}{932191}
\definecolor{dkpink}{rgb}{1.2,0,1.6}
\definecolor{iden}{HTML}{0332FF}
\definecolor{comment}{HTML}{B12122}

\algnewcommand\Assert{\textbf{assert} }

\usepackage{listings}

\let\C=\lstinline

\usepackage{etoolbox}
\expandafter\patchcmd\csname \string\lstinline\endcsname{%
  \leavevmode
  \bgroup
}{%
  \leavevmode
  \ifmmode\hbox\fi
  \bgroup
}{}{%
  \typeout{Patching of \string\lstinline\space failed!}%
}

\tikzset{math3d/.style=
    {x= {(-0.353cm,-0.353cm)}, z={(0cm,1cm)},y={(1cm,0cm)}}}

\title{A Formal Disproof of the Hirsch Conjecture}

\author{Xavier Allamigeon}
\orcid{0000-0002-0258-8018}
\affiliation{\institution{Inria}\institution{CMAP, CNRS, {\'E}cole polytechnique, Institut Polytechnique de Paris}\country{France}}
\email{xavier.allamigeon@inria.fr}
\author{Quentin Canu}
\orcid{0000-0001-8461-2458}
\affiliation{\institution{Inria}\institution{CMAP, CNRS, {\'E}cole polytechnique, Institut Polytechnique de Paris}
\country{France}}
\email{quentin.canu@inria.fr}
\author{Pierre-Yves Strub}
\orcid{0000-0002-8196-7875}
\affiliation{\institution{Meta}\country{France}}
\email{strubpy@meta.com}

\begin{abstract}
The purpose of this paper is the formal verification of a counterexample of Santos et al.\ to the so-called Hirsch Conjecture on the diameter of polytopes (bounded convex polyhedra). In contrast with the pen-and-paper proof, our approach is entirely computational: we implement in \Coq{} and prove correct an algorithm that explicitly computes, within the proof assistant, vertex-edge graphs of polytopes as well as their diameter. The originality of this certificate-based algorithm is to achieve a tradeoff between simplicity and efficiency. 

Simplicity is crucial in obtaining the proof of correctness of the algorithm. This proof splits into the correctness of an abstract algorithm stated over proof-oriented data types and the correspondence with a low-level implementation over computation-oriented data types. A special effort has been made to reduce the algorithm to a small sequence of elementary operations (\eg{}, matrix multiplications, basic routines on sets and graphs), in order to make the derivation of the correctness of the low-level implementation more transparent.

Efficiency allows us to scale up to polytopes with a challenging combinatorics. For instance, we formally check the two counterexamples of Matschke, Santos and Weibel to the Hirsch conjecture, respectively $20$- and $23$-dimensional polytopes with $\numprint{36425}$ and $\numprint{73224}$ vertices involving rational coefficients with up to $40$ digits in their numerator and denominator. We also illustrate the performance of the method by computing the list of vertices or the diameter of well-known classes of polytopes, such as (polars of) cyclic polytopes involved in McMullen's Upper Bound Theorem.
\end{abstract}

\keywords{polyhedra, polytopes, Hirsch Conjecture, proof assistants, certified computation}

\begin{document}
\maketitle 
\thispagestyle{empty}
\section{Introduction}

\subsection{Motivations}

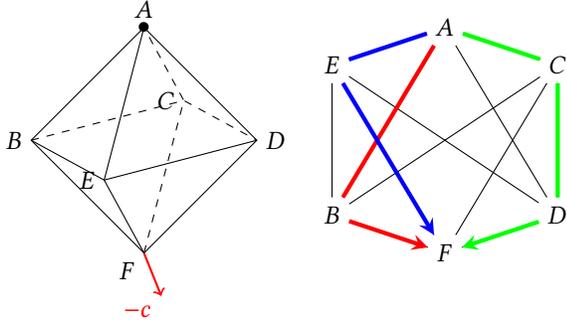
\begin{figure}[t]
\begin{tikzpicture} [math3d]
\draw (0,0,-1.5) -- (0,-1.5,0) -- (0,0,1.5) -- (0,1.5,0) -- cycle;
\draw (0,0,1.5) -- (1.5,0,0) -- (0,0,-1.5);
\draw (0,-1.5,0) -- (1.5,0,0) -- (0,1.5,0);
\draw[dashed] (0,-1.5,0) -- (-1.5,0,0) -- (0,1.5,0);
\draw[dashed] (0,0,-1.5) -- (-1.5,0,0) -- (0,0,1.5);
\draw[thick, ->, red] (0,0,-1.5) -- (0.2,0.3,-2) node[below left] {$-c$};

\draw (0,0,1.5) node {$\bullet$};
\draw (0,0,1.5) node[above] {$A$};
\draw (0,-1.5,0) node[left] {$B$};
\draw (-1.5,0,0) node[left] {$C$};
\draw (0,1.5,0) node[right] {$D$};
\draw (1.5,0,0) node[left] {$E$};
\draw (0,0,-1.5) node[below left] {$F$};

\node (B) at (0,4,-1.5) {$F$};
\node (L) at (0,2.5,-1) {$B$}; 
\node (T) at (0,4,1.5) {$A$};
\node (R) at (0,5.5,-1) {$D$};
\node (F) at (0,2.5,1) {$E$};
\node (Bc) at (0,5.5,1) {$C$};

\draw  (L) -- (T) -- (R) -- (B) -- (L);
\draw (L) -- (F) -- (R) -- (Bc) -- (L);
\draw (T) -- (F) -- (B) -- (Bc) -- (T);
\draw[ultra thick, ->, >=stealth, red] (T) -- (L) -- (B);
\draw[ultra thick, ->, >=stealth, blue] (T) -- (F) -- (B);
\draw[ultra thick, ->, >=stealth, green] (T) -- (Bc) -- (R) -- (B);

\end{tikzpicture}
\caption{Several possible execution traces of the simplex method on the $3$-dimensional cross polytope (see \Cref{sec:experiments} for a description). The execution of the simplex method draws a path in the graph of the polytope from a starting vertex to an optimal one (rightmost part of the picture).}\label{fig:simplex}
\end{figure}

The study of diameters of polyhedra is at the heart of the following major problem in optimization: \emph{does the simplex method terminate in polynomial time?} This question is open since Georg~B.~Dantzig introduced the simplex method in the late 40s, and it has inspired to the Fields medalist Steve Smale the ninth of his problems for the 21\textsuperscript{st} century on the existence of a strongly polynomial algorithm for linear programming~\cite{Smale98}. The simplex method~\cite{Dantzig51} is certainly the most standard technique to solve \emph{linear programs}, \ie, problems of the form
\[
\begin{array}{r@{\quad}l}
\Minimize & \scalar{c}{x} \\[\jot]
\SubjectTo & \scalar{a_1}{x} \geq b_1 \, , \; \dots \, , \; \scalar{a_m}{x} \geq b_m \, , \\[\jot] & x \in \R^n
\end{array}
\]
for some vectors $a_1, \dots, a_m, c \in \R^n$ and reals $b_1, \dots, b_m$, where $\scalar{y}{z} \coloneqq \sum_{i = 1}^n y_i z_i$ denotes the Euclidean scalar product of $y, z \in \R^n$. It consists in minimizing the objective function $x \mapsto \scalar{c}{x}$ over the convex polyhedron formed by the points $x \in \R^n$ satisfying the constraints $\scalar{a_i}{x} \geq b_i$ for all $i = 1, \dots, m$.
From a geometric perspective, the principle of the simplex method is to iteratively decrease the objective function by visiting a subset of vertices of the polyhedron. More precisely, at every iteration, the method selects a vertex with smaller value among the vertices which are adjacent (\ie, connected by an edge) to the current vertex. We refer to \Cref{sec:preliminaries} for the mathematical definitions of vertices and edges of polyhedra, and to \Cref{fig:simplex} for an illustration. The choice of the next vertex at every iteration is specified by the so-called \emph{pivot rule}. In consequence, every pivot rule makes the simplex method draw a path in the \emph{vertex-edge graph} of the polyhedron, \ie, the graph defined by the vertices and edges of the polyhedron (see \cref{fig:simplex}). While a large number of pivot rules have been described in the literature, all the rules that have been mathematically analyzed have been shown to exhibit (sub)exponential behavior in the worst case (see~\cite{AmentaZiegler99}, and~\cite{Disser22} for a more recent account).

Recall that, in a graph, the \emph{distance} between two vertices is the length (\ie, the number of edges) of any shortest path between them. The \emph{diameter} of the graph is then defined as the largest distance between any two vertices. Consequently, the \emph{(combinatorial) diameter} of a polyhedron, \ie, the diameter of its graph, constitutes a lower bound on the number of iterations performed by the simplex method with any pivot rule. With this motivation, Hirsch formulated the following conjecture in a letter to Dantzig in 1957:
\begin{conjecture}[Hirsch conjecture over polytopes]
The diameter of any $d$-dimensional polytope with $p$ facets is bounded by $p - d$.
\end{conjecture}
In this statement, a \emph{polytope} refers to the convex hull of finitely many points, or equivalently, a bounded polyhedron. The \emph{facets} of a $d$-dimensional polyhedron are the faces of dimension $d-1$. (Originally, the conjecture was stated over polyhedra rather than polytopes, but it was quickly realized that the bound does not hold over unbounded polyhedra~\cite{KleeWalkup67}.) The study of the diameter of polytopes and polyhedra has received a tremendous attention over the years; see~e.g.~\cite{survey_ziegler} for a survey on the topic. Still, the conjecture remained unsolved for more than fifty years, until Santos exhibited a counterexample:
\begin{theorem}[\cite{Santos2012}]
There exists a $43$-dimensional polytope with $86$ facets and diameter larger than $43$.
\end{theorem}
In a further work joint with Matschke and Weibel~\cite{Matschke2015}, Santos provided two other counterexamples to the conjecture, respectively in dimension $20$ and $23$. All these constructions critically rely on a computational argument. More precisely, in all of them, the non-Hirsch polytope is obtained from a smaller dimensional polytope with a special combinatorial structure, that of a \emph{spindle}, and in which the distance between two distinguished vertices has to be proved larger than the dimension. For the original counterexample, this spindle has dimension $5$, $48$ facets and $322$ vertices. Santos indicates that the property has been checked thanks to the informal software Polymake~\cite{polymake:2000}, and he provides two independent proofs that are ``computer-free (but not computation-free).'' Fortunately, exploiting the symmetry group of this spindle makes the computations manageable by hand.

Since Santos' breakthrough, the interest for diameters of polytopes and polyhedra has remained intact. First, this is still not understood to which extent a counterexample to the Hirsch conjecture can be found in smaller dimensions. Moreover, from the perspective of the complexity of the simplex method, the most important question is whether or not the diameter of polyhedra can be bounded by a polynomial in the dimension and the number of facets; this is known as the \emph{polynomial Hirsch conjecture}. It is likely that computations will play a key role in any progress on these two questions, just like they did in Santos' construction. 

In consequence, there is a strong motivation to develop a framework in which computations over polyhedra are performed in a proof assistant. This would considerably enlarge the scope of research for pathological polytopes (\eg, giving the capacity to scale up to larger numbers of vertices) while retaining (if not increasing) the level of trust compared to pen-and-paper computations. Combinatorial properties of polytopes are not the only topic such a contribution would benefit to. For instance, polyhedra are a central tool in critical applications such as software compilation or verification~\cite{CousotHalbwachs78}, or invariant computation in the analysis of dynamical systems~\cite{Guglielmi2017}. In these applications, the computation of the vertices of a polyhedron is a core primitive, and developing a formally proved algorithm that performs this operation is a noteworthy goal.

\subsection{Contributions}

We introduce an algorithm which computes the set of vertices as well as the graph of polytopes, and we present its implementation and proof of correctness in the proof assistant~\Coq{}. The originality of our contribution lies in the combination of simplicity of design and practical efficiency. Simplicity is the key feature which makes the algorithm realistically implementable \emph{and} provable in a proof assistant. Simplicity also has major advantages in terms of maintainability, and portability to other proof assistants.

Our algorithm is based on certificates, \ie{}, it takes as input a graph  computed by some informal software, and formally certifies that this graph is indeed correct. In more details, the certificate is the graph of a perturbation of the polytope that has remarkable properties such as being connected and regular. Thanks to these properties, the certification consists of a sequence of elementary steps, involving basic operations such as matrix multiplications, set operations, \etc{}. The graph of the original polytope can be then deduced as the image of the former graph by some mapping. 

The algorithm is first implemented and proved correct by using some proof-oriented types, \eg{}, dependent types representing matrices, convex polyhedra, \etc. However, these types would have a prohibitive cost for computations: they have been defined to ease the formal development of mathematical theories, prioritizing the use of naive data-structures and algorithms over computationally well-behaving ones.
As a consequence, we implement a second algorithm over computation-oriented types, \eg{}, persistent arrays, native integers, arbitrary precision rationals. An additional benefit of the elementary structure of the algorithm is to make the equivalence proof between the low-level and high-level implementations easier.

In order to demonstrate the practical efficiency of our approach, we make experiments on several classes of polytopes of interest. We manage to compute the graph of the $20$- and $23$-dimensional counterexamples of~\cite{Matschke2015} to the Hirsch conjecture, respectively with $\numprint{36425}$ and $\numprint{73224}$ vertices, and $\numprint{364250}$ and $\numprint{842076}$ edges, and we deduce a formal disproof of the conjecture. We also study classic polytopes such as hypercubes, cross-polytopes, and polars of cyclic polytopes. The latter are known to maximize the number of vertices for fixed dimension and number of facets, as stated by McMullen's Upper Bound Theorem~\cite{McMullen70}.

\subsection{Related Work}

Computing the vertices of polytopes and polyhedra is a notoriously difficult problem. As the number $v$ of vertices can be exponential in the number $m$ of defining inequalities and the dimension $n$, this complexity has to be measured as a function of $m$, $n$ and $v$. The worst-case complexity of the problem is not fully understood: it is an $\mathsf{NP}$-hard enumeration problem for unbounded polyhedra~\cite{Khachiyan2008}, but its status is still open for polytopes. The two main vertex enumeration algorithms are the double description method~\cite{MRTT53,FukudaProdon96} and the reverse search method~\cite{Avis1992,Avis2000}. While the complexity of the double description method cannot be easily bounded in terms of $m$, $n$ and $v$, the complexity of the reverse search method is in $O(\textrm{poly}(m,n) \mathinner{v})$ for ``nondegenerate'' representations of polytopes (we refer to~\Cref{sec:simple_algo} for the definition). This falls in the same complexity class as our algorithm, since the two approaches rely on the enumeration of simplex bases. The double description method and the reverse search method have standard implementations~\cite{cddlib,lrslib} that are widely used in software dedicated to computational mathematics~\cite{polymake:2000, sage} as well as software verification~\cite{PPL}. 

Computational proofs, or proofs by reflection, is nowadays a common technique in the \Coq{} proof assistant. For instance, it is used in the implementation of tactics that rely on symbolic computations~\cite{gregoire:hal-00819484}, and has been widely used for the formal proof of the \emph{four color theorem}~\cite{10.1007/978-3-540-87827-8_28}. Proofs by reflection is also at the root of the \emph{small scale reflection} (SSReflect~\cite{gonthier:inria-00515548}) proof methodology, where one makes a pervasive use of computation for solving goals that involve decidable predicates. We refer to~\cite{Hendriks02proofreflection,DBLP:conf/sfp/JamesH09,braibant:hal-00383070,chlipala_certified_2013} for other examples of computational based proofs in the \Coq{} proof assistant. Proof by reflection has also been used in many other systems: Agda~\cite{vanderwalt:hal-00987610}, Isabelle/HOL~\cite{DBLP:journals/jar/ChaiebN08}, Lean~\cite{leanrefl}, PVS~\cite{10.1007/BFb0055152}. In relation with polyhedral computations, Farkas' certification techniques motivated by application to static analysis have appeared in ~\cite{Fouilhe2014,Boulme2018}. In contrast with our work, this only cover polyhedral computations using inequalities, and not the computation of vertices or vertex-edge graphs. We finally mention that an implementation of a simplex-based satisfiability procedure has been done by~\cite{Spasic2012} in the proof assistant Isabelle.

Last, our paper makes use of \emph{program} and \emph{data structure} refinement techniques, a proof methodology that consists in transforming a high-level program or data structure to a lower-level, more efficient one, while preserving the main properties (program specification and/or data structure invariants). Refinement techniques have already been used in the context of the \Coq{} proof assistant, notably in \CoqEAL{}~\cite{DenesMortbergSiles12,Cohen2013}, a \Coq{} framework for easing the definition of data structure refinements. Our formalization of data refinements closely follows the approach of \CoqEAL{}. The difference is that, for a first experiment, we have chosen not to exploit the automated deduction of equivalence proofs provided by \CoqEAL{} (based on \Coq{} typeclasses). In this way, we have a finer-grained control of the low-level implementation of our algorithm. We point out that the burden of proving the equivalence proof by hand was limited thanks to the simplicity of our algorithm.

\subsection{Organization of the Paper}

In \Cref{sec:preliminaries}, we recall basic notions on polyhedra and introduce notation. In Section~\ref{sec:simple_algo}, we define the certificate-based algorithm computing the graph of a given polytope. \Cref{sec:Implementation} is dedicated to the implementation of the latter algorithm using computation-oriented data types, and deals with the proof of correctness of this implementation. In \Cref{sec:Hirsch}, we bring the formal disproof of the Hirsch Conjecture. Finally, we report on the experiments of our approach on several classes of polytopes in \Cref{sec:experiments}.

The source files of the present submission can be found in the \texttt{git} repository of the \CoqPolyhedra{} library\footnote{\url{https://github.com/Coq-Polyhedra/Coq-Polyhedra/tree/CPP-23}}. We often refer to these source files in the paper. On top of \Coq{}, we rely on the \MathCompShort{} library~\cite{MathComp} as well as the \C$finmap$ and \C$bignums$ libraries.\footnote{\url{https://github.com/math-comp/finmap}}$^{,}$\footnote{\url{https://github.com/coq-community/bignums}}

We note that this work is a (slightly) extended version of the conference paper~\cite{CPP23} published in the proceedings of CPP'23.

\section{Preliminaries and Notation}\label{sec:preliminaries}

As discussed in the introduction, a \emph{(convex) polyhedron} is defined as the set of points $x \in \R^n$ satisfying finitely many affine inequalities $\scalar{a_i}{x} \geq b_i$, where $a_1, \dots, a_m \in \R^n$ and $b_1, \dots, b_m \in \R$. We shall also write the system of constraints as $A x \geq b$, where $A \in \R^{m \times n}$ is the matrix with rows $\trans{a_i}$,  $b = (b_i)_i \in \R^m$, and $\geq$ denotes the entrywise ordering over $\R^m$. A notable subclass of polyhedra are polytopes. A \emph{polytope} is the convex hull $\big\{ \sum_{i = 1}^p \lambda_i v^i \colon \lambda_1, \dots, \lambda_p \geq 0 \, , \; \sum_{i = 1}^p \lambda_i = 1\big\}$ of finitely many points $v^1, \dots, v^p \in \R^n$. Minkowski Theorem states that polytopes are precisely the bounded polyhedra.

\begin{figure}[t]
\begin{tikzpicture} [math3d]
\draw (0,0,1.5) node {$\bullet$};
\draw (0,0,-1.5) -- (0,-1.5,0) -- (0,0,1.5) -- (0,1.5,0);
\draw[ultra thick] (0,1.5,0) -- (0,0,-1.5);
\draw (0,0,1.5) -- (1.5,0,0) -- (0,0,-1.5);
\draw (0,-1.5,0) -- (1.5,0,0) -- (0,1.5,0);
\draw[dashed] (0,-1.5,0) -- (-1.5,0,0) -- (0,1.5,0);
\draw[dashed] (0,0,-1.5) -- (-1.5,0,0) -- (0,0,1.5);
\draw (0,0,1.5) node[above right] {$x$};
\draw (0,0,-1.5) node[below left] {$y$};
\draw (0,1.5,0) node[right] {$z$};
\draw[->] (0,0,1.3) -- (0,0,0.8) node[below] {$c_1$};
\draw[->] (-0.1,0.7,-0.7) -- (-0.6,0.2,-0.2)  node[left]{$c_2$};
\end{tikzpicture}
\caption{The $3$-dimensional cross polytope. The objective vector $c_1$ is minimized by only one vertex $x$, while objective vector $c_2$ is minimized by an edge $[y,z]$ with $y$ and~$z$.}\label{fig:face}
\end{figure}
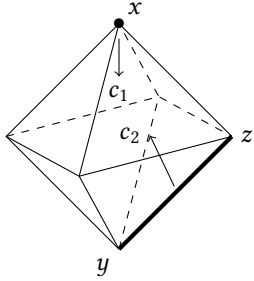

The \emph{dimension} of a polyhedron is defined as the dimension of its \emph{affine hull}, \ie, the smallest (inclusionwise) affine subspace containing the polyhedron. For instance, a point has dimension $0$, a line segment has dimension~$1$, \etc. Let $\Pcal$ be a polyhedron. A (nonempty) \emph{face} of $\Pcal$ is the set of points minimizing some linear function $x \mapsto \scalar{c}{x}$ over $\Pcal$, where $c \in \R^n$. Equivalently, a face is the set of optimal solutions of some linear program over $\Pcal$. The \emph{vertices} and the \emph{edges} of $\Pcal$ are the faces of dimension $0$ and $1$ respectively; see \Cref{fig:face}. Any bounded edge writes as the line segment between two vertices $v, v'$ of $\Pcal$. In this case, the vertices $v$ and $v'$ are said to be \emph{adjacent}. As described in the introduction, the graph of $\Pcal$, denoted by $\Gvert(\Pcal)$ (or simply $\Gvert$ when clear from context), is the combinatorial graph induced by the adjacency relation over the vertices.

In the rest of the paper, we define $[p] \coloneqq \{1, \dots, p\}$ for all integer $p \geq 1$. The cardinality of a finite set $S$ is denoted by~$\# S$. Given a matrix $M \in \R^{p \times q}$, we denote by $M_i$ its $i$th row ($i \in [p]$), and, for all subset $I \subset [p]$, by $M_I \in \R^{\# I \times q}$ the submatrix with rows $M_i$ for $i \in I$. The identity matrix of size $p \times p$ is denoted by $\Id_p$. Given a (nonoriented) graph $G = (V, E)$ (where $E \subset V \times V$) and a vertex $v \in V$, we denote by $N_G(v) \coloneqq \{w \in V \colon (v,w) \in E \}$ the neighborhood of $v$ in $G$, \ie, the set of vertices $w$ adjacent to $v$. Given a function $f \colon V \to V'$, the \emph{image of~$G$ by $f$}, denoted by $f(G)$, is defined as the graph with vertices $f(V) \coloneqq \{f(v) \colon v \in V\}$ and edges $\big\{ (f(v), f(w)) \colon (v,w) \in E \, , \; f(v) \neq f(w) \big\}$.

\section{Graph Certification Algorithm}\label{sec:simple_algo}

\subsection{A First Algorithm for the Nondegenerate Setting}

Our approach heavily relies on the notion of bases and basic points, which are the central ingredients of the simplex method. Let $\Pcal = \{x \in \R^n \colon A x \geq b\}$ a polyhedron, where $A \in \R^{m \times n}$ and $b \in \R^m$. We recall that a \emph{basis} is a subset $I \subset [m]$ of cardinality $n$ such that the submatrix $A_I$ is nonsingular. In this case, the equality system $A_I x = b_I$ has a unique solution $x^I$, which we call the \emph{basic point} associated with $I$. When $x^I$ belongs to the polyhedron $\Pcal$, the basis $I$ is said to be \emph{feasible}. By extension, the point $x^I$ is said to be a \emph{feasible basic point}.

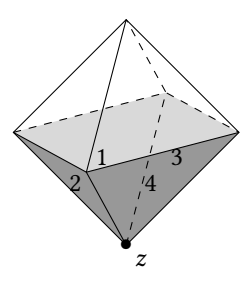
\begin{figure}[t]
\hfill
\adjustbox{valign=c}{\begin{tikzpicture} [math3d]
\draw (0,0,-1.5) -- (0,-1.5,0) -- (0,0,1.5) -- (0,1.5,0) -- cycle;
\draw (0,0,1.5) -- (1.5,0,0) -- (0,0,-1.5);
\draw (0,-1.5,0) -- (1.5,0,0) -- (0,1.5,0);
\draw[dashed] (0,-1.5,0) -- (-1.5,0,0) -- (0,1.5,0);
\draw[dashed] (0,0,-1.5) -- (-1.5,0,0) -- (0,0,1.5);
\fill[opacity=0.3] (0,0,-1.5) -- (0,-1.5,0) -- (1.5,0,0);
\fill[opacity=0.3] (0,0,-1.5) -- (0,1.5,0) -- (1.5,0,0);
\fill[opacity=0.15] (0,0,-1.5) -- (0,-1.5,0) -- (-1.5,0,0);
\fill[opacity=0.15] (0,0,-1.5) -- (0,1.5,0) -- (-1.5,0,0);
\draw (0.5,-0.5,-0.5) node {$2$};
\draw (0.5,0.5,-0.5) node {$4$};
\draw (-0.5,-0.5,-0.5) node {$1$};
\draw (-0.5,0.5,-0.5) node {$3$};
\draw (0,0,-1.5) node {$\bullet$};
\draw (0,0,-1.5) node[below right] {$z$};
\end{tikzpicture}}\hfill
$\left\{\begin{aligned}
&\text{1} & & x_1 + x_2 + x_3 &\geq -1\\
&\text{2} & & - x_1 + x_2 + x_3 &\geq -1\\
&\text{3} & & x_1 - x_2 + x_3 &\geq -1\\
&\text{4} & & - x_1 - x_2 + x_3 &\geq -1\\
&\text{5} & & x_1 + x_2 - x_3 &\geq -1\\
&\text{6} & & - x_1 + x_2 - x_3 &\geq -1\\
&\text{7} & & x_1 - x_2 - x_3 &\geq -1\\
&\text{8} & & - x_1 - x_2 - x_3 &\geq -1\\
\end{aligned}\right.$
\hfill\null

\caption{The $3$-dimensional cross polytope has degenerate bases. For instance, the bottom point $z$ ($x_3 = -1$) is a vertex associated with four feasible bases: $(2,3,4)$, $(1,3,4)$, $(1,2,4)$ and $(1,2,3)$.}\label{fig:degenerate_bases}
\end{figure}

It is a standard property that the vertices of $\Pcal$ are precisely the feasible basic points; see~\cite[Chapter~11]{Schrijver86}. However, in general, the correspondence between them is not bijective: every feasible basic point is a vertex, but a vertex may be the basic point associated with more than one feasible basis; see \Cref{fig:degenerate_bases} for an illustration. These bases are said to be \emph{degenerate}. We say that we are in the \emph{nondegenerate setting} when there are no such bases, \ie, the correspondence between feasible bases and vertices is one-to-one.

Given a linear program of the form 
\begin{equation}\label{eq:lp}
\Minimize \quad \scalar{c}{x} \quad \SubjectTo \quad A x \geq b \, , \; x \in \R^n \, ,
\end{equation}
where $c \in \R^n$, the simplex method iterates over feasible bases up to reaching a (feasible) basic point that minimizes the function $x \mapsto \scalar{c}{x}$. In this scheme, any two consecutive bases $I, I'$ satisfy $\# (I \cap I') = n-1$, \ie, they only differ by one element. Such bases are said to be \emph{adjacent}. This adjacency relation gives rise to the \emph{graph of (feasible) bases}, that we denote by $\Gbases$. The relation between basic points and vertices extends to the graph of the polyhedron and the graph of bases as follows:
\begin{proposition}\label{prop:graph_rel}
The vertex-edge graph $\Gvert$ is the image of the graph $\Gbases$ by the function $I \mapsto x^I$ which maps any feasible basis to its basic point.

Moreover, in the nondegenerate setting, the latter function is an isomorphism between the two graphs.
\end{proposition}
This result elaborates on the geometric description of the simplex that we made in the introduction, \ie, the simplex method induces a path in the vertex-edge graph of the polyhedron. 

In this section, we discuss the computation of the vertex-edge graphs of polytopes in the nondegenerate case only. The exposition of this special case is done to facilitate the understanding of the general algorithm presented in \Cref{subsec:general}. We remark that, unless explicitly stated, we did not have to formalize the results presented below.

In light of the second part of \Cref{prop:graph_rel}, we exploit the isomorphism between $\Gvert$ and $\Gbases$, and we sketch an algorithm certifying that a given graph (computed \emph{a priori} by some informal procedure) coincides with the graph of feasible bases. To this purpose, we exploit the following two fundamental properties:
\begin{proposition}\label{prop:connected}
The graph $\Gbases$ is connected.
\end{proposition}

\begin{proposition}\label{prop:regular}
In the nondegenerate setting, and when $\Pcal$ is a polytope, the graph $\Gbases$ is $n$-regular, \ie, every feasible basis is adjacent to precisely $n$ feasible bases.
\end{proposition}

Proposition~\ref{prop:connected} holds in a general setting (even with degenerate bases). It is a consequence of the fact that, for any feasible basis $I^\star$, we can find a vector $c \in \R^n$ such that $x^{I^\star}$ is the unique optimal solution of the linear program~\eqref{eq:lp}. Then, the simplex method initialized with any feasible basis~$I$ draws a path to $I^\star$ in the graph $\Gbases$. 

\begin{algorithm}[t]
\caption{Graph certification algorithm (nondegenerate setting and $\Pcal = \{x \in \R^n \colon A x \geq b \}$ bounded)}
\label{algo:simple_algo}
\begin{algorithmic}[1]
\Require $A \in \R^{m \times n}$, $b \in \R^m$ and $G = (V,E)$
\State \Assert $G$ is nonempty\label{line:nonempty}
\sForAll{$I \in V$} \label{line:feasible} \Assert $I$ is a feasible basis
\ForAll{$I \in V$}
    \sForAll{$J \in N_G(I)$} \Assert $\#(I \cap J) = n - 1$\label{line:adjacent}
    \State \Assert $\# N_G(I) = n$ \label{line:card}
\EndFor
\State \Return $\True$
\end{algorithmic}
\end{algorithm}

The algorithm covering the nondegenerate setting and the assumption that $\Pcal$ is a polytope is \Cref{algo:simple_algo}. We use the command \textbf{assert} <\textit{cond}> as some syntactic sugar for the block \textbf{if not} <\textit{cond}> \textbf{then} \textbf{return} $\False$. The vertices of the input graph $G = (V, E)$ are supposed to be sets $I$ of integers. The algorithm consists in four steps: 
\begin{enumerate}[(i)] 
\item check that the graph is nonempty (Line~\lineref{line:nonempty});
\item check that every vertex $I$ is a feasible basis (Line~\lineref{line:feasible});
\item for each vertex $I \in V$, check that its neighborhood consists of adjacent bases (Line~\lineref{line:adjacent});
\item for each vertex $I \in V$, check that its neighborhood has cardinality $n$ (Line~\lineref{line:card}).
\end{enumerate} 
The second and third steps ensures that $G$ is a subgraph of $\Gbases$. Moreover, as $\Gbases$ is $n$-regular (\Cref{prop:regular}), the fourth step actually ensures that $N_G(I) = N_{\Gbases}(I)$ for all $I \in V$. In consequence, $G$ is a subgraph of $\Gbases$ such that the neighborhood of every vertex in $G$ agrees with that in $\Gbases$. As shown in the following lemma, the nonemptiness of $G$ and the connectedness of $\Gbases$ then ensure that $G$ and $\Gbases$ are identical:
\begin{lemma} \label{lemma:subgraph iso}
Let $G$ and $H$ two graphs such that $G$ is a nonempty subgraph of $H$. Suppose that $H$ is connected, and $N_G(v) = N_H(v)$ for all vertices $v$ of $G$. Then $G = H$.
\end{lemma}
This lemma is given by \C$Lemma |*sub_gisof*|$ (see \C$high_graph.v$) in the source of the project. (This result actually shows a slightly more general though equivalent statement, where $G$ is replaced by an isomorphic graph.) The next result then follows from \Cref{lemma:subgraph iso}, and shows that \Cref{algo:simple_algo} is correct: 
\begin{theorem}\label{th:simple_algo_correct}
Suppose that we are in the nondegenerate setting, and that $\Pcal = \{x \in \R^n \colon A x \geq b \}$ is a polytope. If \Cref{algo:simple_algo} returns $\True$, then $G = \Gbases$.
\end{theorem}
The reader can also verify that if one of the assertions in  \Cref{algo:simple_algo} fails, then the graph $G$ cannot be equal to $\Gbases$. 

\subsection{Dealing with the General Case}\label{subsec:general}

\subsubsection{A Perturbation Approach} 

\begin{figure}[t]
\begin{tikzpicture}[math3d]
\coordinate (A) at (0.31,-0.95,0);
\coordinate (B) at (1,0,0);
\coordinate (C) at (0.31,0.95,0);
\coordinate (D) at (-0.81,0.59,0);
\coordinate (E) at (-0.81,-0.59,0);
\coordinate (F) at (0,0,2);
\draw (A) -- (B) -- (C) -- (D);
\draw[dashed] (A) -- (E) -- (D);
\draw (A) -- (F);
\draw (B) -- (F);
\draw (C) -- (F);
\draw (D) -- (F);
\draw[dashed] (E) -- (F);
\draw (F) node {$\bullet$};
\coordinate (A') at (0.31,3-0.95,0);
\coordinate (B') at (1,3,0);
\coordinate (C') at (0.6,3+0.55,0);
\coordinate (D') at (-0.81,3+0.09,0);
\coordinate (E') at (-0.81,3-0.59,0);
\coordinate (F'1) at (-0.34,3-0.25,1.16);
\coordinate (F'2) at (0.08,3-0.25,1.48);
\coordinate (F'3) at (0.42,3,1.16);
\draw (A') -- (B') -- (C') -- (D');
\draw[dashed] (A') -- (E') -- (D');
\draw[dashed] (F'1) -- (E');
\draw (F'1) -- (D');
\draw (F'1) -- (F'2);
\draw (F'2) -- (A');
\draw (F'2) -- (F'3);
\draw (F'3) -- (B');
\draw (F'3) -- (C');
\draw (F'1) node {$\bullet$};
\draw (F'2) node {$\bullet$};
\draw (F'3) node {$\bullet$};
\end{tikzpicture}
\caption{The lexicographic perturbation approach. The top vertex on the left-hand side is split into three distinct vertices on the right-hand side.}\label{fig:lex_perturbation}
\end{figure}
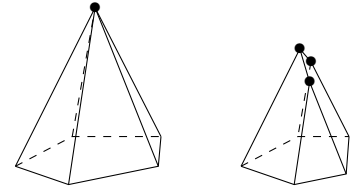

The nondegenerate setting does not hold in general, but we can reduce to it by slightly perturbing the polytope, while still keeping a way to recover the main combinatorial structure such as the set of vertices or the vertex-edge graph. Our approach originates from the so-called lexicographic pivot rule introduced by Dantzig~et~al.~\cite{Dantzig1955}, and later exploited by Avis~\cite{Avis2000} for his vertex enumeration algorithm~\cite{lrslib}. It consists in perturbing the vector $b$ by replacing each entry $b_i$ ($i \in [m]$) by the quantity $b_i - \eps^i$, where $\eps > 0$ is a sufficiently small real. Geometrically, while the normal vectors to the hyperplanes delimiting the polyhedron are still the same, the perturbation of the vector $b$ breaks every vertex associated with several (degenerate) bases into distinct (and nondegenerate) basic points; see \Cref{fig:lex_perturbation} for an illustration.

Instead of instantiating $\eps$ by some numerical values (which would raise the problem of determining how small it should be), the perturbation is achieved in a symbolic way, by thinking of each~$b_i$ as a polynomial in $\eps$ of degree at most $m$. This leads to considering a ``polyhedron'' where the entries of the points are polynomials in $\eps$ of degree at most $m$ as well. Such polynomials $\sum_{k = 0}^m \alpha_k \eps^k$ can be encoded as row vectors $(\alpha_0, \dots, \alpha_m)$ of size $1+m$. In this case, the standard order over the reals is replaced by the lexicographic order $\leqlex$ over $(1+m)$-tuples. Indeed, we have $(\alpha_0, \dots, \alpha_m) \leqlex (\beta_0, \dots, \beta_m)$ if and only if $\sum_{k = 0}^m \alpha_k \eps^k \leq \sum_{i = 0}^m \beta_k \eps^k$ for all $0 < \eps \ll 1$. This gives rise to the \emph{symbolically perturbed polyhedron}
\begin{equation}\label{eq:Ptilde}
\Ptilde \coloneqq \big\{\,X \in \R^{n \times (1+m)} \colon A X \geqlex \btilde\,\big\}
\end{equation}
where the matrix $\btilde \coloneqq \begin{bmatrix} b & -\Id_m \end{bmatrix} \in \R^{m \times (1+m)}$ corresponds to the perturbation of the vector $b$ described above: the $i$th row of $\btilde$ encodes the polynomial $b_i - \eps^i$. In~\eqref{eq:Ptilde}, the relation $\geqlex$ stands for the entrywise extension of the lexicographic order: two matrices $X, Y \in \R^{p \times (1+m)}$ satisfies $X \geqlex Y$ if $X_i \geqlex Y_i$ for all $i \in [p]$. The matrices $X \in \R^{n \times (1+m)}$ in $\Ptilde$ correspond to vectors with perturbed entries  (encoded as polynomials in $\eps$ of degree at most $m$).

The notion of feasible bases still makes sense in case of such symbolically polyhedra. To avoid any confusion with the bases of the original polyhedron, the feasible bases of $\Ptilde$ are referred to as lex-feasible bases. Formally, a set $I \subset [m]$ is a \emph{lex-feasible basis} if $I$ has cardinality $n$, the matrix $A_I$ is nonsingular, and the basic point $X^I \coloneqq A_I^{-1} \btilde_I$ satisfies $A X^I \geqlex \btilde$. We recall that lex-feasible bases are not degenerate:
\begin{lemma}[see~\cite{AllamigeonKatzJAR2018} for a formalization] 
Let $I, I'$ be two distinct lex-feasible bases. Then $X^I \neq X^{I'}$. 
\end{lemma}
 
\subsubsection{Formalizing the Properties of the Lex-Graph}\label{subsec:formalization}

In what follows, we assume that $\Pcal$ is a polytope.

The adjacency relation defined over feasible bases carries over to lex-feasible bases in a straightforward way. This induces the \emph{graph of lex-feasible bases}, or \emph{lex-graph} for short, that we denote by $\Glex$.

Lex-feasible bases form the cornerstone of the formalization of the simplex method done by Allamigeon and Katz~\cite{AllamigeonKatzJAR2018}, and provided in the library \CoqPolyhedra{}. In more details, the latter work formalized the \emph{lex-simplex method}, which iterates over lex-feasible bases in order to avoid cycling on degenerate bases. \CoqPolyhedra{} introduces the type \C$lex_feasible_basis A b$ of lex-feasible bases, where \C$A : 'M_(m,n)$ and \C$b : 'cV_m$ correspond to the matrix~$A$ and the (unperturbed) vector $b$. (We recall that \C$'M_(m,n)$ and \C$'cV_m$ are respectively the types of $\C$m$ \times \C$n$ $-matrices and \C$m$-vectors provided by the library \MathCompShort{}~\cite{MathComp}.) We build on this and start by defining the graph of lex-feasible bases, denoted \C$lex_graph$:
\begin{lstlisting}
Definition |*set_adjacence*| := ((*(* adjacency relation *)*))
  fun I I' : {set 'I_m} =>  #| I :&: I' | == n.-1.§\smallskip§
Definition |*lex_graph*| := mk_graph 
  [fset x | x : Simplex.lex_feasible_basis A b] set_adjacence.
\end{lstlisting}
We then prove that, as expected, the properties of \Cref{prop:connected,prop:regular} hold in the case of lex-feasible bases:
\begin{lstlisting}
Lemma |*lex_graph_connected*| : connected lex_graph.§\smallskip§
Lemma |*lex_graph_regular*| : regular lex_graph n.
\end{lstlisting}
Their proofs are straightforward consequences of the formalization of the lex-simplex method provided in \CoqPolyhedra{}.

One fundamental property of lex-feasible bases is that they constitute a subset of the feasible bases of $\Pcal$. In more details, we denote by $\pi$ the function which maps a matrix $X \in \R^{n \times (1+m)}$ (\ie, a perturbed point) to its first column. The latter corresponds to the unperturbed part of $X$, \ie, the value of the perturbed point when $\eps = 0$. The following result is folklore (we refer to~\cite{AllamigeonKatzJAR2018} for the formalization): 
\begin{proposition}\label{prop:lex_feasible_basis_subset}
Let $I$ be a lex-feasible basis. Then $I$ is a feasible basis of $\Pcal$, and $\pi(X^I)$ is the associated basic point.
\end{proposition}

We extend the latter result to the following correspondence between the lex-graph and the vertex-edge graph.
\begin{theorem}\label{th:im_lex_graph}
The vertex-edge graph $\Gvert$ of $\Pcal$ is the image of $\Glex$ by the function $\phi \colon I \mapsto \pi(X^I)$. 
\end{theorem}
In \Coq{}, this statement is written as follows (see Module \C$enum_proof.v$):
\begin{lstlisting}
Theorem |*im_lex_graph_vert_graph*| :
  poly_graph P = 
    (Simplex.point_of_basis b) @/ lex_graph.
\end{lstlisting}
where the term \C$poly_graph P$ is the vertex-edge graph $\Gvert$, the function \C$Simplex.point_of_basis b$ corresponds to the function $\phi$, and a term of the form \C$f @/ G$ stands for the image of a graph \C$G$ by the function \C$f$.

We briefly comment on the formal proof of \Cref{th:im_lex_graph} since, most often, its informal proof is not detailed in the literature. The hardest part of the proof is to show that for every edge $[v,w]$ of $\Pcal$, there exist two adjacent lex-feasible bases $I, J$ such that $v = \phi(I)$ and $w = \phi(J)$. We construct these two bases by exploiting the lex-simplex method of \CoqPolyhedra{}. More precisely, since $[v,w]$ is an edge of $\Pcal$, there exists a vector $c$ such that $[v,w]$ is precisely the set of points of $\Pcal$ minimizing the function $x \mapsto \scalar{c}{x}$. Calling the lex-simplex method with the objective vector~$c$ provides a lex-feasible basis $I_0$ such that the point $\phi(I_0)$ reaches this minimal value. Since $\phi(I_0)$ is a vertex of $\Pcal$, this should be either $v$ or $w$. Without loss of generality, we assume that $\phi(I_0) = v$. We now consider an objective vector $c'$ such that $w$ is the only point minimizing $x \mapsto \scalar{c'}{x}$ over $\Pcal$ (this is possible by definition of a vertex), and introduce a third objective vector $d \coloneqq c + \delta c'$, where $\delta > 0$ is a sufficiently small quantity. Intuitively, perturbing $c$ into $d$ in this way should ensure that $w$ is the unique minimizer of $x \mapsto \scalar{d}{x}$ over $\Pcal$, and that $v$ is the second ``best'' vertex after $w$, \ie, $\scalar{d}{w} < \scalar{d}{v} < \scalar{d}{z}$ for every vertex $z \notin \{v,w\}$.\footnote{These relations actually imply how small $\delta$ needs to be chosen.} We finally apply the lex-simplex method with objective function $x \mapsto \scalar{d}{x}$, starting from the lex-feasible basis $I_0$. Since the objective function cannot increase along the way, the lex-simplex method generates a sequence $I_0, \dots, I_{p-1}, I_p, \dots$ of adjacent lex-feasible bases such that $\phi(I_k) = v$ for all $k < p$ and $\phi(I_p) = w$. Then, it suffices to take $I = I_{p-1}$ and $J = I_p$.

\subsubsection{Certification of the Lex-Graph and the Vertex-Edge Graph}\label{subsubsec:cert}

In light of the properties formalized in \C$Lemma lex_graph_connected$ and \C$Lemma lex_graph_regular$, we can derive from \Cref{algo:simple_algo} a method certifying that an informally computed graph coincides with the lex-graph. Then, \Cref{th:im_lex_graph} will allow us to recover the vertex-edge graph of $\Pcal$ from the latter by computing its image by $\phi$. 

We describe the formalization of the certification procedure for $\Glex$. It takes as input a graph \C$G$ whose vertices are pairs of the form \C$(I, X)$, where \C$I : {set 'I_m}$ is a subset of integers (less than $m$), and \C$X : 'M_(n,1+m)$ is a $(n \times (1+m))$-matrix. These two components are intended to represent a lex-feasible basis~$I$ and the corresponding basic point $X^I$ respectively. The algorithm is formalized as a program returning a Boolean value (reminding that \C$&&$, \C$[&& ...]$ and \C$[forall ..., ...]$ stand for Boolean conjunctions):
\begin{lstlisting}
Definition |*high_enum_algo*| G : bool :=
  (G != graph0) ((*(* G is nonempty *)*)) &&
  [forall u : vertices G, ((*(* u is a pair (I,X) *)*))
    [&& card_verification u, 
        bas_verification u, 
        feas_verification u,
        reg_verification u & 
        subset_verification u 
    ]
  ].
\end{lstlisting}
where we define
\begin{lstlisting}
Definition |*card_verification*| u := #|u.1| == n.
Definition |*bas_verification*| u := 
  (row_submx A u.1) *m u.2 == row_submx b_pert u.1.
Definition |*feas_verification*| u := 
  u.2 \in Simplex.lex_polyhedron A b_pert.
Definition |*reg_verification*| u := 
  #|` successors G u| == n.
Definition |*subset_verification*| u := 
  [forall u' : successors G u, 
     set_adjacence u.1 u'.1].
\end{lstlisting}
We recall that \C$u.1 : T$ and \C$u.2 : T'$ respectively correspond to the first and second components of a pair \C$u : T * T'$. The implementation of \C$high_enum_algo$ follows the structure of \Cref{algo:simple_algo}. In particular, it starts by checking that the graph \C$G$ is nonempty, and then performs five consecutive tests on every vertex \C$u$. The last two tests apply to the neighborhood \C$successors G u$ of \C$u$ in \C$G$, and respectively check that it has cardinality $n$ (\C$reg_verification$) and consists of adjacent bases (\C$subset_verification$). The main difference with \Cref{algo:simple_algo} is the way the first component $I$ of every vertex  \C$u$ in \C$G$ is verified to be a lex-feasible basis. This is the purpose of the first three tests \C$card_verification$, \C$bas_verification$ and \C$feas_verification$. The first one checks that $I$ has cardinality $n$, while the last two ones respectively make sure that $A_I X = \btilde_I$ and $A X \geqlex \btilde$. We note that, since $\btilde = \begin{bmatrix} b & -\Id_m \end{bmatrix}$, the equality $A_I X = \btilde_I$ ensures that $A_I$ is a nonsingular matrix. Indeed, if $Y$ is the submatrix of $X$ formed by its $(1+i)$th columns for $i \in I$, it can be verified that $A_I Y = -\Id_n$. In other words, the matrix $X$ already carries a certificate that $A_I$ is nonsingular. As a consequence, $I$ is a basis. Since $A_I X = \btilde_I$ then $X = X^I$, and the condition $A X \geqlex \btilde$ finally ensures that $I$ is a lex-feasible basis. 

This is how we arrive at the proof of the correctness of \C$high_enum_algo$:
\begin{lstlisting}
Theorem |*repr_lex_graph*| G : 
  high_enum_algo G -> 
    gisof G lex_graph (fun u => u.1).
\end{lstlisting}
The \C$gisof$ predicate in the rightmost part of the implication means that the function $(I, X) \mapsto I$ is an isomorphism between \C$G$ and the lex-graph.

The following statement is obtained by combining \C$Theorem im_lex_graph_vertex_graph$ and \C$Theorem repr_lex_graph$:
\begin{lstlisting}
Theorem |*repr_poly_graph*| G :
  high_enum_algo G -> poly_graph P = (phi @/ G).
\end{lstlisting}
where the function \C$phi$ is defined as \C$phi u = col 0 u.2$, \ie, if \C$u = (I, X)$, then \C$phi u$ is the first column of the matrix \C$X$.

\section{Efficient Implementation} \label{sec:Implementation}

The \Coq{} function \C$high_enum_algo$ introduced in \Cref{sec:simple_algo} works with dependent types (\eg, \MathCompShort{} types) that are adapted for proof but not for computation. For example, natural numbers are expressed in unary form, and rationals carry proof terms of the fact that their numerator and denominator are coprime. Similarly, the implementations of finite sets, graphs or matrices are based on \MathCompShort{} sequences (\ie, basic lists built by induction) that are not made for fast computations, and are provided with multiple proof terms for their well-formedness. As a consequence, the function \C$high_enum_algo$ cannot return within a reasonable amount of time even on the simplest instances. To overcome this problem, we exploit data types that are closer to machine representations and thus practically more efficient. Based on these, we implement a ``low-level'' version of the function \C$high_enum_algo$; see \Cref{subsec:low-level-data}. In \Cref{subsec:refinements}, we describe how we relate high-level data structures with low-level ones by combining refinements. Finally, in \Cref{subsec:proof_equivalence}, we deal with the proof of equivalence of the low-level implementation of the function \C$high_enum_algo$.

\subsection{Low-Level Implementation}\label{subsec:low-level-data}

The main data types used in the low-level implementation of the certification algorithm are the following: 
\begin{inparaenum}[(i)]
\item the type \C$int$ of 63-bits integers (module \C$Int63$ in \Coq{}) built on OCaml integers;
\item the type \C$array$ of persistent arrays (module \C$PArray$ in \Coq{}) built on OCaml arrays and based on the ideas of~\cite[Section~2]{CF07};
\item the type \C$bigQ$ that represents arbitrarily large rationals (module \C$BigQ$ of \Coq{}~library \C$bignums$) built on sequences of words based on 63-bits integers~\cite{GreT06}.
\end{inparaenum}

The type \C$bigQ$ is useful to manipulate polyhedra in which the numerical entries of the inequality system or of the vertices can be very large rationals, as in the counterexamples to the Hirsch conjectures that we deal with in \Cref{sec:Hirsch}. As we describe next, persistent arrays are involved in various data structures in the low-level implementation. We point out that we choose persistent arrays over \Coq{} AVL trees, because our early experiments have shown that the latter suffer performance issues. Indeed, every tree comes with a proof term for balancing, and this term can grow excessively on large instances. 

Vectors and matrices with rational entries are implemented using the types \C$array bigQ$ and \C$array (array bigQ)$ respectively. A system of inequalities defining a polyhedron is then represented by a term of type \C$polyType := array (array bigQ) * array bigQ$. Bases, which are sets of row indices, are encoded with the type \C$array int$. More precisely, the elements of a basis are collected in a sorted array. This allows us to compute the intersection of two bases in linear time.

Graphs whose vertices are labeled with some type \C$t$, \ie, the low-level counterpart of graphs of type \C$graph t$, are implemented using pairs of the form \C$(g, lbl)$, where \C$g : array (array int)$ and \C$lbl : array t$. In more details, the vertices of a low-level graph are indexed by integers (of type \C$int$), and the term \C$lbl.[i]$ corresponds to the label of the vertex of index \C$i$. The term \C$g$ represents the adjacency array of the graph, \ie, \C$g.[i]$ is the array containing the indices of the neighbors of the vertex of index \C$i$. We use the notation \C$graph_struct := array (array int)$. The indexing of vertices is made in such a way that the labels in the array \C$lbl$ are sorted in nondecreasing order (to this extent, we introduce a total order relation over labels). This ensures that every label occurs only once in the graph.

Using these data structures, we build a low-level implementation, called \C$enum_algo$, of the function \C$high_enum_algo$; see Module \C$enum_algo.v$. The function \C$enum_algo$ is a transparent adaptation of \C$high_enum_algo$ on low-level data structures. In particular, every test in \C$high_enum_algo$ has a counterpart on the low-level side. Unlike high-level data structures based on dependent types, our low-level data structures do not come with well-formedness invariants for free. Instead, these invariants have to be checked by extra functions. For instance, we need to verify that arrays representing vectors and matrices have a specific size (\eg, the dimension of the ambient space, or the number of inequalities), that all the vertex indices appearing in a graph belong to the right range, and that arrays representing sets (such as bases or graph labels) are sorted. The advantage of our certification algorithm is that we do not need to prove that such invariants are preserved throughout the function. Indeed, owing to the simplicity of the algorithm, data structures are only accessed for reading, and no new structure is produced. In consequence, checking the consistency of data structures occurs only once, before the call to the function \C$enum_algo$.

Once \C$enum_algo$ has verified that a low-level graph is the (low-level representation of) the lex-graph of the polytope, it remains to deal with the low-level computation of the image of the lex-graph by the function \C$phi$ defined in \C$Theorem repr_poly_graph$ (\cf~end of \Cref{subsubsec:cert}). This is the final step to get the vertex-edge graph of the polytope. While the image of a graph is a basic construction for high-level graphs, the problem is slightly more involved on the low-level side. Once again, we rely on certificates. In more details, we define a function \C$img_lex_graph$ that takes as input two graphs \C$g_lex g_vert : graph_struct$ and their respective labelings \C$lbl_lex$ and \C$lbl_vert$, and checks that \C$(g_vert, lbl_vert)$ is the image of \C$(g_lex, lbl_lex)$ by the low-level counterpart \C$low_phi := fun u => (u.2).[0]$ of the function \C$phi$. We give a short description of the implementation of this function.

Additional certificates \C$morph$, \C$morph_inv$ and \C$edge_inv$ are provided to the function \C$img_lex_graph$. The term \C$morph : array int$ corresponds to a mapping between the indices of the vertices of the two graphs. The function \C$img_lex_graph$ first checks that this mapping is consistent with the function \C$low_phi$ over labels, in the sense that, for every index \C$i$ of \C$g_lex$, \C$low_phi lbl_lex.[i]$ is equal to \C$lbl_vert.[morph.[i]]$. The remaining part of \C$img_lex_graph$ consists in verifying that \C$morph$ induces a graph morphism between the two incidence structures described by \C$g_lex$ and \C$g_vert$. To this purpose, the algorithm checks that \C$morph.[i] < length g_vert$ for all indices \C$i$, which ensures \C$morph$ to be well-formed. Then, it makes sure that the mapping \C$morph$ is surjective. This is done by exploiting the certificate \C$morph_inv : array int$ intended to be a right-inverse of \C$morph$, and by checking that for all indices \C$i$ of \C$g_vert$, we have \C$morph.[morph_inv.[i]] == i$. The algorithm then proceed with edges. It checks that if \C$i j : int$ are adjacent in \C$g_lex$, then \C$morph.[i]$ and \C$morph.[j]$ are adjacent in \C$g_vert$ as well, unless \C$morph.[i] == morph.[j]$ (by definition of the image). Conversely, it exploits the third certificate \C$edge_inv$ to verify for any two adjacent vertices \C$i'$, \C$j'$ of \C$g_vert$, there exist two adjacent vertices \C$i$, \C$j$ of \C$g_lex$ such that \C$morph.[i'] == i$ and \C$morph.[j'] == j$.\footnote{We warn that the preimages of \C$i'$ and \C$j'$ provided by \C$morph_inv$ may not be adjacent in \C$g_lex$, which is why we need the additional certificate \C$edge_inv$.}

\subsection{Data Refinements}\label{subsec:refinements}

In order to prove that the low-level implementations are correct w.r.t.~the high-level ones, we follow the approach introduced in the project~\CoqEAL{}~\cite{DenesMortbergSiles12,Cohen2013} and use data refinements. Given a high-level type \C$T$ and the corresponding low-level type \C$t$, a \emph{refinement} \C$r : t -> T -> Prop$ is a relation between terms that respectively correspond to the low-level and the high-level representation of a same object. In this setting, two functions are equivalent if they return related outputs given related inputs. This is formalized as follows:
\begin{lstlisting}
Definition |*rel_func*|
  (r1 : t -> T -> Prop) (r2 : u -> U -> Prop)
  (f : t -> u) (g : T -> U) := 
  (forall x y, r1 x y -> r2 (f x) (g y)).
\end{lstlisting}
We use the notation \C$(r1 =~> r2) f g$ for \C$rel_func r1 r2 f g$. 

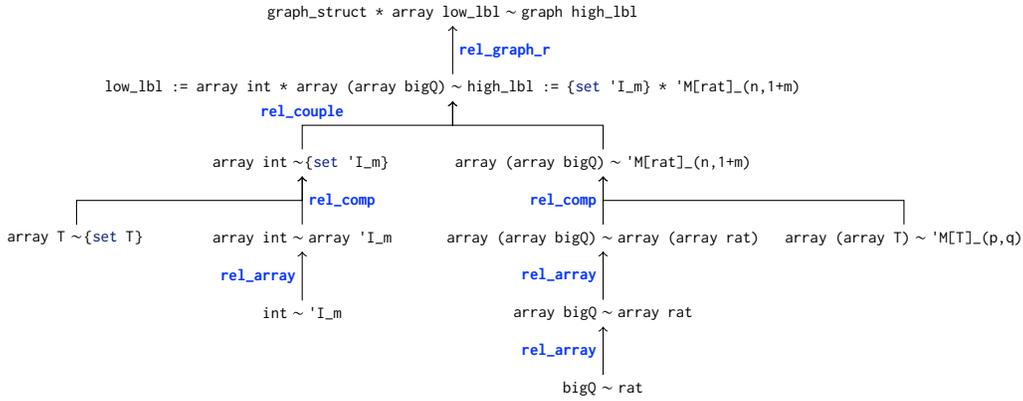
\begin{figure*}[t]
\begin{center}
\begin{tikzpicture}[every node/.style={scale=0.7}]
\node (BQR) at (0,0) {\C$bigQ$${}\sim{}$\C$rat$};
\node (arrBQR) at (0,1) {\C$array bigQ$ $\sim$ \C$array rat$};
\node (matBQR) at (0,2) {\C$array (array bigQ)$ $\sim$ \C$array (array rat)$};
\node (matrat) at (4,2) {\C$array (array T)$ $\sim$ \C$'M[T]_(p,q)$};
\node (point) at (0,3) {\C$array (array bigQ)$ $\sim$ \C$'M[rat]_(n,1+m)$};
\draw[->] (BQR) -- (arrBQR) node[left,midway] {\C$|*rel_array*|$};
\draw[->] (arrBQR) -- (matBQR) node[left,midway] {\C$|*rel_array*|$};
\draw[->] (matrat) |- (0,2.5) -| (point);
\draw[->] (matBQR) -- (point) node[left,midway] {\C$|*rel_comp*|$};

\node (IO) at (-4,1) {\C$int$ $\sim$ \C$'I_m$};
\node (arrIO) at (-4,2) {\C$array int$ $\sim$ \C$array 'I_m$};
\node (setord) at (-7,2) {\C$array T$${}\sim{}$\C${set T}$};
\node (basis) at (-4,3) {\C$array int$ $\sim$ \C${set 'I_m}$};
\draw[->] (IO) -- (arrIO) node[left,midway] {\C$|*rel_array*|$};
\draw[->] (arrIO) -- (basis) node[right,midway] {\C$|*rel_comp*|$};
\draw[->] (setord) |- (-4,2.5) -| (basis);

\node (couple) at (-2,4) {\C$low_lbl := array int * array (array bigQ)$ $\sim$ \C$high_lbl := {set 'I_m} * 'M[rat]_(n,1+m)$};
\draw[->] (basis) |- (-2,3.5) node[midway,above]{\C$|*rel_couple*|$} -| (couple);
\draw[->] (point) |- (-2,3.5) -| (couple);

\node (final) at (-2,5) {\C$graph_struct * array low_lbl$ $\sim$ \C$graph high_lbl$};
\draw[->] (couple) -- (final) node[right,midway] {\C$|*rel_graph_r*|$};
\end{tikzpicture}
\end{center}
\caption{Construction of the refinement relation \C$rel_lex_graph$ between the low-level and high-level representations of lex-graphs. We make use of the notation \C$t$ $\sim$ \C$T$ to state that there is a refinement relation between the low-level type \C$t$ and the high-level type \C$T$.}\label{fig:refinement}
\end{figure*}

The tree in \Cref{fig:refinement} describes the combination of refinement relations performed in order to relate low-level and high-level representations of lex-graphs, \ie, graphs over pairs of bases and matrices. (The formalization of the relations is carried out in the module \C$refinement.v$.) We comment on this tree.

The leaves of the tree correspond to refinement relations between atomic types. For every such relation, we need to prove that the implementations of basic operations over the low-level and the high-level types are equivalent. For instance, the refinement \C$rel_int_ord : int -> 'I_m -> Prop$ relates computationally efficient integers with \MathCompShort{} ordinals of type \C$'I_m$ (\ie, unary integers less than a fixed integer \C$m$), and we prove the equivalence of between the low-level and high-level ordering. This states as: 
\begin{lstlisting}
Lemma |*rel_int_ord_lt*| : 
  (rel_int_ord =~> rel_int_ord =~> eq) 
    (fun i j : int => (i <? j)%int63) 
    (fun i j : 'I_m => (i < j)%nat)
\end{lstlisting}
where \C$eq$ is the identity relation. Other basic refinements include the relation \C$rel_array_set : array T -> {set T} -> Prop$ between the implementation of finite sets with arrays and the corresponding \MathCompShort{} type \C${set T}$, or the relation \C$rel_mx_col : array (array T) -> 'M[T]_(p,q) -> Prop$ between array-based matrices and \MathCompShort{} matrices of size $\C$p$\times \C$q$ $. Finally, we use the refinement relation \C$rat_bigQ : bigQ -> rat -> Prop$ implemented in \CoqEAL{} between the type \C$bigQ$ of computationally efficient rationals and the \MathCompShort{} type \C$rat$ of rationals.

The edges of the tree correspond to \emph{functors}. The latter allow us to combine refinements in order to construct more complex relations. For example, the functor \C$rel_array$ lifts a refinement \C$r : t -> T -> Prop$ to another between the types \C$array t$ and \C$array T$:
\begin{lstlisting}
Definition |*rel_array*| r a A := 
  length a = length A 
    /\ forall i, i < length a -> r a.[i] A.[i].
\end{lstlisting}
It comes with the proof of equivalence for several basic primitives, \eg, computing the image \C$arr_map f a$ of some array \C$a$ by some function \C$f$:
\begin{lstlisting}
Lemma |*rel_array_map*| r r' f F :
  (r =~> r') f F -> 
    (rel_array r =~> rel_array r') 
       (arr_map f) (arr_map F).
\end{lstlisting}
Similarly, the functor \C$rel_couple$ combine two refinement relations \C$r : t -> T -> Prop$ and \C$r' : t' -> T' -> Prop$ into a refinement relation between the product types \C$t * t'$ and \C$T * T'$. A third functor \C$rel_comp$ allows us to compose a refinement relation between \C$t$ and \C$T$ from two refinement relations \C$r : t -> tT -> Prop$ and \C$r' : tT -> T -> Prop$. For example, the relation between the type \C$array (array bigQ)$ of low-level rational matrices and the type \C$'M[rat]_n$ of high-level square matrices is provided by composing the relation between low-level matrices of \C$bigQ$ and \C$rat$ with the relation \C$rel_mx_col$ previously discussed instanced with \C$T := rat$. 

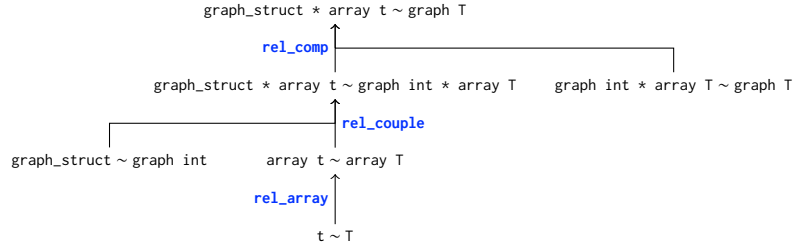
\begin{figure*}[t]
\begin{center}
\begin{tikzpicture}[every node/.style={scale=0.7}]
\node (couple) at (-2,4) {\C$t$ $\sim$ \C$T$};
\node (arrt) at (-2,5) {\C$array t$ $\sim$ \C$array T$};
\node (struct) at (-5,5) {\C$graph_struct$ $\sim$ \C$graph int$};
\node (lblgraph) at (-2,6) {\C$graph_struct * array t$ $\sim$ \C$graph int * array T$};
\node (gisof) at (2.5,6) {\C$graph int * array T$ $\sim$ \C$graph T$};
\node (final) at (-2,7) {\C$graph_struct * array t$ $\sim$ \C$graph T$};
\draw[->] (couple) -- (arrt) node[midway, left] {\C$|*rel_array*|$};
\draw[->] (arrt) -- (lblgraph) node[midway, right] {\C$|*rel_couple*|$};
\draw[->] (struct) |- (-2,5.5) -| (lblgraph);
\draw[->] (lblgraph) -- (final) node[midway,left] {\C$|*rel_comp*|$};
\draw[->] (gisof) |- (-2,6.5) -| (final);
\end{tikzpicture}
\end{center}
\caption{Construction of the functor \C$rel_graph_r$ between low-level and high-level graphs, from a refinement relation \C$t$${}\sim{}$\C$T$.}\label{fig:rel_graph_r}
\end{figure*}

As shown in \Cref{fig:refinement}, the refinement relation between low-level and high-level lex-graphs is obtained by applying the functor \C$rel_graph_r$ to the refinement relation between the types \C$low_lbl$ and \C$high_lbl$ (that respectively correspond to the low-level and high-level representations of lex-graph labels). The functor \C$rel_graph_r$ is described in \Cref{fig:rel_graph_r}. Given a relation between a low-level type \C$t$ and a high-level type \C$T$, it builds a refinement relation between the type \C$graph_struct * array t$ of low-level graphs labeled by \C$t$ and high-level graphs labeled by \C$T$. It is built by applying the functors previously described to two atomic refinement relations. The first one, called \C$rel_structure$, relates the low-level type \C$graph_struct := array (array int)$ of incidence arrays to the type \C$graph int$ high-level graphs over (low-level) integers. The second one relates the type \C$graph int * array T$, which corresponds to a mixed representation of graphs in which the incidence structure is stored separately from the mapping to labels, to the type \C$graph T$ of high-level graphs.

\subsection{Proof of Equivalence}\label{subsec:proof_equivalence}

We are now ready to prove the equivalence between the low-level and high-level implementations of the lex-graph certification algorithm, \ie, \C$enum_algo$ and \C$high_enum_algo$ respectively; see Module \C$enum_equiv.v$. It involves the refinement relation \C$rel_poly$ between low-level and high-level representations of the inequality system defining the polyhedron (between \C$polyType$ and the type \C$'M[rat]_(m,n) * 'cV[rat]_m$ of pairs \C$(A,b)$), and the refinement relation \C$rel_lex_graph$ between low-level and high-level lex-graphs described in \Cref{subsec:refinements}. The equivalence is stated as follows:
\begin{lstlisting}
Lemma |*lex_certif_equiv*| :
  (rel_poly =~> rel_lex_graph =~> eq)
    enum_algo high_enum_algo.
\end{lstlisting}
If we put aside straightforward program transformations, the proof essentially consists in unfolding the definition of every test in the functions \C$enum_algo$ and \C$high_enum_algo$, and to use the equivalence between the basic operations involved on the low-level and high-level sides, such as lexicographic comparison of row vectors, matrix-vector multiplication, or set intersection.

Analogously, the correctness of the low-level function \C$img_lex_graph$ certifying the image of the lex-graph by the function \C$phi$ is given by the following statement:
\begin{lstlisting}
Lemma |*img_lex_graph_equiv*| (...) :
  rel_lex_graph (g_lex, lbl_lex) G_lex ->
  rel_vert_graph (g_vert, lbl_vert) G_vert ->
  img_lex_graph morph morph_inv edge_inv 
    g_lex lbl_lex g_vert lbl_vert ->
  G_vert = phi @/ G_lex.
\end{lstlisting}
where \C$rel_vert_graph$ is an instantiation of the \C$rel_graph_r$ functor for graphs labelled with vectors. (We point out that this statement is not an equivalence because of the additional certificates \C$morph$, \C$morph_inv$ and \C$edge_inv$ provided to \C$img_lex_graph$.)

In practice, the certificates provided to the algorithms are written using low-level types. Indeed, as explained in \Cref{sec:experiments}, these certificates can be very large, and it would impossible to compile them using high-level types. In order to exploit the correctness statements that we previously discussed, every refinement relation \C$rel_t_T : t -> T -> Prop$ comes with an extra function \C$spec_t_T : t -> T$ which builds a high-level term from a well-formed low-level one, so that we have \C$r x (spec_t_T x)$. In this way, we arrive at the following statement, which 
\begin{lstlisting}
Theorem |*Validation*| Po (...) :
  (...) 
  enum_algo Po g_lex lbl_lex ->
  img_lex_graph morph morph_inv edge_inv 
    g_lex lbl_lex g_vert lbl_vert ->
  poly_graph (poly_of_syst (spec_poly Po)) =
    spec_vert_graph (g_vert, lbl_vert).
\end{lstlisting}
It takes as input the low-level certificates as well as the hypotheses that they are certified by the low-level implementation of the algorithms. It proves that, for the high-level polyhedron \C$poly_of_syst (spec_poly Po)$ described by the low-level inequality system \C$Po$, the (high-level) vertex-edge graph corresponds to the low-level graph \C$(g_vert, lbl_vert)$. The proof is a combination of the correctness of the high-level algorithm (\C$Theorem repr_poly_graph$) with that of the low-level implementations (\ie, \C$Lemma lex_certif_equiv$ and \C$Lemma img_lex_graph_equiv$).

\section{Formal Disproof of Hirsch Conjecture} \label{sec:Hirsch}

The purpose of this section is to describe how we arrive at the formal disproof of the Hirsch Conjecture using the certification algorithms and their low-level implementations described in \Cref{sec:Implementation}. Recall that the Hirsch conjecture makes use of three notions: the diameter of the vertex-edge graph, the number of facets, and the dimension of the polytope. We disprove the conjecture by computing lower bounds on the diameter and the dimension and an upper bound on the number of facets, to conclude by transitivity. We first explain how we deal with the computation of each quantity. 

Considering a polyhedron $\Pcal = \{x \in \R^n \colon A x \geq b \}$ where $A \in \R^{m \times n}$ and $b \in \R^m$, it is well known that the number of facets is bounded by the number $m$ of inequalities. This statement has been added to the properties of polyhedra established in \CoqPolyhedra{} (see Module \C$poly_base.v$), and is expressed as follows:
\begin{lstlisting}
Lemma |*facets_le*| {base : base_t[R,n]}:
  (#|` facets 'P(base) | <= #|` base|)%nat.
\end{lstlisting}
The term \C$base$ correspond to a set of inequalities defining the polyhedron, and \C$'P(base)$ is the term representing the polyhedron. 

To deal with the dimension, we establish that the (informal)  counterexamples to the Hirsch conjecture have dimension equal to $n$ (\ie, the dimension of the ambient space). This is achieved by exhibiting a set of $n+1$ vertices $v^0, \dots, v^n \in \R^n$ of the polytope that are \emph{affinely independent}, \ie, the matrix $M = \begin{bmatrix} v^1 - v^0 & \dots & v^n - v^0 \end{bmatrix}$ is nonsingular. These points are provided by means of (low-level) certificates \C$map_lbl : array int$  and \C$origin : int$ which represent the set of their indices in the low-level vertex-edge graph, along with an informally computed low-level matrix \C$inv_lbl$ that we verify to be the inverse of the matrix $M$. On the low-level side, the verification of these certificates is performed by the function \C$dim_full_test$ (in \C$enum_algo.v$), and the correctness statement is expressed as follows (see \C$enum_high_algo.v$):
\begin{lstlisting}
Lemma |*high_dim_h*| : 
  dim_full_test map_lbl origin inv_lbl -> 
  \pdim P = n.+1.
\end{lstlisting}
where \C$\pdim P$ stands for the dimension of the polytope \C$P$ shifted by one (\CoqPolyhedra{} uses the convention that the emptyset have dimension $0$, points have dimension $1$, \etc). 

Finally, the lower bound on the diameter is computed by providing a vertex $v$ of the graph whose \emph{eccentricity}, \ie, the maximal distance between $v$ and any other vertex $w$, reaches the value of the diameter. In order to compute the eccentricity, we exploit the Breadth-First Search (BFS) algorithm, and implement it on low-level and high-level data structures in the module \C$diameter.v$. The low-level implementation is provided by the function \C$Low.BFS : graph_struct -> int -> NArith.BinNat$ which takes as input a low-level incidence array and the index of a vertex, and returns the maximal length of shortest paths from it. Similarly, the high-level implementation is the function \C$High.BFS$ working with the type \C$graph T$. The correspondence with the low-level implementation is expressed as follows:
\begin{lstlisting}
Lemma |*rel_struct_BFS*| (g : graph_struct) (G : graph int): 
  rel_structure g G -> 
    forall x, mem_vertex g x->
      Low.BFS g x = High.BFS G x :> nat.
\end{lstlisting}

We are now ready to formally disprove the conjecture, by running our certification algorithms on the two counterexamples to the Hirsch conjecture provided in~\cite{Matschke2015}. This is done in the directories \C$test/data/poly20dim21$ and \C$test/data/poly23dim24$ respectively (we refer to the \C$README$ file to extract them from the zipped archives in \C$test/archives$). We exploit the explicit inequality representations of the counterexamples provided by Weibel.\footnote{\url{https://sites.google.com/site/christopheweibel/research/hirsch-conjecture}} They take the form of input files for the library \C$lrslib$~\cite{lrslib}. We use this library together with additional Python scripts to generate the informal certificates for our algorithms (they are stored in \C$test/data/polyXXdimYY/coq/$). As explained in \Cref{subsec:proof_equivalence}, the latter are provided using low-level types. We give a short description of every certificate:
\begin{itemize}
\item \C$poly.v$ contains the description of the polytope by inequalities;
\item \C$g_lex.v$ and \C$lbl_lex.v$ (resp.\ \C$g_vert.v$ and \C$lbl_vert.v$) are intended to represent the lex-graph (resp.\ the vertex-edge graph);
\item \C$morph.v$, \C$morph_inv.v$ and \C$edge_inv.v$ are the certificates provided to the function \C$img_lex_graph$ (see \Cref{subsec:low-level-data});
\item \C$map_lbl.v$, \C$origin.v$ and \C$inv_lbl.v$ are the certificates for the dimension required by \C$dim_full_test$;
\item \C$cert.v$ provides certificates to check that the input polyhedron is a polytope. These certificates are used to show that nonnegative combinations of the inequalities defining the polyhedron yield inequalities of the form $-K \leq x_i \leq K$ for all $i \in [n]$, where $K$ is a sufficiently large constant. The verification is performed by the function \C$bounded_Po_test$ (module \C$enum_algo.v$).
\item \C$start.v$ is the index of vertex that is used to get a lower bound on the diameter.
\end{itemize}
Each file is compiled using \C$coqc$, and then imported in order to get the formal disproof of the conjecture in \C$test/data/polyXXdimYY/coq_Hirsch/Hirsch.v$:
\begin{lstlisting}
Theorem |*Hirsch_was_wrong*| : 
  exists (d : nat) (P : 'poly[rat]_d),
    (High.diameter (poly_graph P) 
      > #|`facets P| - (\pdim P).-1)%nat.
Proof.
pose P := poly_of_syst (A, b)).
exists n'.+1, P.
apply/disprove_Hirsch.
- exact: well_formedness_ok.
- exact: enum_algo_ok.
- exact: img_graph_ok.
- exact: bounded_Po_test_ok.
- exact: dim_full_test_ok.
- exact: diameter_check_ok.
Qed.
\end{lstlisting}
(Note that the statement use \C$(pdim P).-1$ for the dimension because of the shift-by-one convention in \CoqPolyhedra{}.) The proof starts by exhibiting the witness of the existential statements, \ie, the polyhedron provided in \C$poly.v$. It then applies \C$Theorem |*disprove_Hirsch*|$ which establishes that the conjecture does not hold provided that all certification tests return \C$true$ (see file \C$theories/enum_equiv.v$). These latter hypotheses are then verified in the last six lines of the proof. They respectively correspond to the verification of 
\begin{inparaenum}[(i)]
\item the well-formedness of all input certificates,
\item the lex-graph,
\item the vertex-edge graph,
\item the boundedness of the polyhedron,
\item the dimension of the polyhedron,
\item the lower bound on the diameter.
\end{inparaenum}
Every test is achieved by using the tactic \C$vm_compute$ of \Coq{}, which computes the (Boolean) result of each test. 

As explained in the introduction, the approach of~\cite{Santos2012}, also used in~\cite{Matschke2015}, builds non-Hirsch polytopes by lifting special low-dimensional spindles to higher dimension. This only provides a lower bound on the diameter of the non-Hirsch polytopes; see~\cite[Th.~1.5]{Santos2012}. In addition to verifying the lower bound, we formally certify the exact value of the diameter of the $20$- and $23$-dimensional counterexamples of~\cite{Matschke2015}:
\begin{lstlisting}
Theorem |*poly20dim21_diameter*| :
  diameter (poly_graph poly20dim21) = 21%nat.§\smallskip§
Theorem |*poly23dim24_diameter*| :
  diameter (poly_graph poly23dim24) = 24%nat.
\end{lstlisting}
Since the two polytopes respectively have $40$ and $46$ facets, this entails that their diameter matches the lower bound. As far as we know, this is the first proof of this fact.

\section{Practical Experiments}\label{sec:experiments}

For the sake of reproducibility, experiments have been conducted on two different architectures: \begin{inparaenum}[(i)] \item a machine with an Apple M1 processor and 32\,GB~RAM running Mac~OS~11.0.1 (Architecture~A), 
\item a machine with a 2.3\,GHz Intel Core processor and 64\,GB RAM running Linux~5.4 (Architecture~B).
\end{inparaenum}
Both use Coq~8.16.1.

As described in the introduction, the two counterexamples of~\cite{Matschke2015} are very involved polytopes, with several tens of thousands vertices and hundreds of thousands edges. Moreover, the vertices have long rational coefficients, with up to 40 digits long numerator and denominator. As a consequence, explicitly writing the certificates in plain files lead to very large \C$.v$ files whose compilation is a challenging (if not unrealistic) task for \Coq{}. For instance, for the $23$-dimensional counterexample, we need about 600\,MB to store the term \C$lbl_lex$ containing the labels of the lex-graph in such a plain \C$.v$ file. In this case, the memory used by \C$coqc$ during the compilation of the file turns out to be the main limiting factor. The compilation runs out of memory and fails with Architecture~B. Thanks to a better memory management (memory compression and use of swap memory), it succeeds on Architecture~A, but it takes more than $\numprint{8000}$\,s to complete. To overcome these compilation issues, we have implemented a \Coq{} plugin called \C$BinReader$\footnote{\url{https://github.com/Coq-Polyhedra/coq-binreader}} that provides a command \C$LoadData <filename.bin> As <term_id>$ building a \Coq{} term \C$<term_id>$ from a description read in the binary file \C$<filename.bin>$. The plugin handles all possible combinations of the data types described in \Cref{subsec:low-level-data}. In practice, the time needed to build the term becomes negligible, and the main consuming step is the typechecking of the term by the kernel of \Coq{}. As an illustration, we manage to build the term \C$lbl_lex$ previously discussed in about 8\,min on Architecture~A and 13\,min on Architecture~B, with a total memory usage limited to~50\,GB.

\Cref{table:experiments} provides the total CPU time spent on each instance using Architecture~A (Architecture~B is around twice slower). The instances \C$cube-n$ and \C$cross-n$ respectively correspond to the $n$-dimensional cubes and cross polytopes. The instances \C$cyclic-p-n$ are the polars of cyclic polytopes in dimension $n$ defined by $p$ facets; see below for the formal description. The instances \C$spindle-n$ correspond to the original small-dimensional spindles used in~\cite{Santos2012, Matschke2015} to build the larger-dimensional counterexamples to the Hirsch conjecture. Finally, \C$poly20dim21$ and \C$poly23dim24$ are respectively the $20$- and $23$-dimensional counterexamples to the Hirsch conjecture given in~\cite{Matschke2015}. The second column of \Cref{table:experiments} provides the time taken by \C$coqc$ to load the certificates using the plugin \C$BinReader$ (which includes the time needed to build the terms from the binary files, and the time used by the kernel to typecheck them). The third column provides the time to check the certificates using the algorithms of \Cref{sec:Implementation}. The last column provides the time to compute the exact diameter of the polytope in \Coq{}. For the instances \C$polyXXdimYY$, we also provide in italics the time taken to compute a lower bound on the diameter, as described in \Cref{sec:Hirsch}. We point out that, in practice, compilation time can be reduced using parallelization (\eg{}, as provided by the build system \C$dune$), as certificate checking and diameter computation can be done  independently (the computation of the diameter is performed on the informal graph).

As we already mentioned, the computation of the results of the certificate tests are carried out using the tactic \C$vm_compute$ in \Coq{}~\cite{GregoireLeroy02}.\footnote{More precisely, we use the tactic \C$vm_cast_no_check$ which relies on \C$vm_compute$, but does the computation only when the kernel typechecks the proof term, \ie, during the execution of the command \C$Qed$.}  The later relies on an optimized bytecode based virtual machine that performs better than the abstract reduction machinery of the \Coq{} kernel when it comes to the full evaluation of computationally intensive function applications. We did not manage to use the more efficient tactic \C$native_compute$~\cite{native-compute} (which consists in the extraction to some OCaml file, its compilation and native execution). Indeed, this tactic failed because of the large size of the terms, and raised some exception. Using \C$vm_compute$, the formal disproof of the Hirsch conjecture takes about 1\,h~22\,min with the $20$-dimensional counterexample, and about 1\,h 55\,min with the $23$-dimensional counterexample. Since these are relatively short times, we have reproduced the same experiments by replacing \C$vm_compute$ by the standard call-by-value conversion tactic \C$cbv$ of \Coq{}. We manage to formally verify the two counterexamples, with approximately a factor 20 slowdown.

\begin{table}
\caption{Total CPU time (in seconds) of the main steps of the certification process on Architecture~A.}\label{table:experiments}
\begin{tabular}{@{}M{1.9cm}M{1.8cm}M{1.5cm}M{1.9cm}@{}}\toprule
instance & certificate loading & certificate checking & diameter computation\\
\midrule
\C$cube_2$ &\tablenum{7.85}&\tablenum{6.40}&\tablenum{1.07} \\
\C$cube_3$ &\tablenum{7.90}&\tablenum{6.30}&\tablenum{1.08} \\
\C$cube_4$ &\tablenum{7.93}&\tablenum{6.31}&\tablenum{1.09} \\
\C$cube_5$ &\tablenum{9.18}&\tablenum{6.31}&\tablenum{1.16} \\
\C$cube_6$ &\tablenum{8.32}&\tablenum{6.36}&\tablenum{1.15} \\
\C$cube_7$ &\tablenum{8.33}&\tablenum{6.45}&\tablenum{1.18} \\
\midrule
\C$cross_2$ &\tablenum{8.54}&\tablenum{6.27}&\tablenum{1.21} \\
\C$cross_3$ &\tablenum{8.11}&\tablenum{6.73}&\tablenum{1.10} \\
\C$cross_4$ &\tablenum{8.17}&\tablenum{6.36}&\tablenum{1.14} \\
\C$cross_5$ &\tablenum{8.65}&\tablenum{6.99}&\tablenum{1.11} \\
\C$cross_6$ &\tablenum{14.97}&\tablenum{17.33}&\tablenum{1.19} \\
\C$cross_7$ &\tablenum{113.32}&\tablenum{265.15}&\tablenum{1.14} \\
\midrule
\C$cyclic_12_6$ &\tablenum{8.37}&\tablenum{6.52}&\tablenum{1.22} \\
\C$cyclic_20_10$ &\tablenum{25.71}&\tablenum{45.87}&\tablenum{36.77} \\
\midrule
\C$spindle_25$ &\tablenum{8.93}&\tablenum{7.06}&\tablenum{1.16} \\
\C$spindle_28$ &\tablenum{10.74}&\tablenum{6.90}&\tablenum{1.21} \\
\C$spindle_48$ &\tablenum{10.88}&\tablenum{8.70}&\tablenum{1.22} \\
\midrule
\C$poly20dim21$ &\tablenum{404.01}&\tablenum{4490.89}&\tablenum{5295.91} \itshape(17.57) \\
\C$poly23dim24$ &\tablenum{1147.00}&\tablenum{5736.19}&\tablenum{24551.46} \itshape(14.75) \\
\bottomrule
\end{tabular}
\end{table}

We finally provide the mathematical descriptions of other instances. The $n$-dimensional cube consists of the points $x \in \R^n$ satisfying $-1 \leq x_i \leq 1$ for all $i \in [n]$. It has $2^n$ vertices. The $n$-dimensional cross-polytope is the set of points $x \in \R^n$ satisfying $\sum_{i = 1}^n |x_i| \leq 1$. It can be described by $2^n$ inequalities of the form $\sum_{i = 1}^n \pm x_i \geq -1$. The challenge is that it has many degenerate bases: every vertex is associated with $2^{n-1}$ feasible bases. This important degeneracy is reflected in the time required to compute the lex-graph. Given $p \geq 1$, the polar of the cyclic polytope, denoted $C(n,p)$, is built by picking $p$ real values $t_1 < \dots < t_p$, and by considering the inequalities $\scalar{c^i - \bar c}{x} \leq 1$ for all $i \in [p]$, where $c^i \coloneqq (t_i, t_i^2, \dots, t_i^n)$ and $\bar c \coloneqq \frac{1}{p} \sum_{i = 1}^p c^i$. McMullen's upper bound theorem~\cite{McMullen70} states that the polytope $C(n,p)$ maximizes the number of vertices among the $n$-dimensional polytope with $p$ facets. This number is given explicitly by:
\[
\binom{p - \ceil{\frac{n}{2}}}{\floor{\frac{n}{2}}} + \binom{p - 1 - \ceil{\frac{n-1}{2}}}{\floor{\frac{n-1}{2}}}
\]
We have checked that the certified graph has precisely the expected number of vertices.

\section{Conclusion}

We have developed practically efficient algorithms that allow us to compute the vertex-edge graphs of polytopes by checking the correctness of informally computed certificates. Simplicity of design is a key feature to reach this goal. We have brought a formal disproof of the well-known Hirsch conjecture, by formally verifying the two counterexamples of~\cite{Matschke2015}. Despite their very involved combinatorial structure, our approach manages to achieve the disproof in a few hours within the proof assistant~\Coq{}. As far as we know, this is the first work showing that computationally demanding proofs on polyhedra can be practically carried out in a proof assistant.

As a future work, we aim at significantly improving the performance of our approach and get much closer to that of informal software. To this purpose, we plan to exploit rank-one updates of matrices and integer pivoting (see~\cite{Avis2000} and the references therein) in order to reduce the size of certificates and improve the computational complexity of matrix-vector multiplications in our algorithms. We also hope that computationally intensive approaches like the one presented in this work will draw more attention to the computational performance of proof assistants. To this extent, we aim at exploiting the more efficient tactic \C$native_compute$~\cite{native-compute} once issues on handling large terms are solved. For the time being, our approach deals only with polytopes. This restriction was made for simplicity, and we plan to extend our techniques by taking into account the unbounded rays of polyhedra in the adjacency graphs. Finally, we plan to investigate to which extent automatic deduction techniques of equivalence proof like the one described in~\cite{Cohen2013} can help to improve the connection between low-level and high-level implementations of our algorithms. 

\begin{acks}
The first author warmly thanks Ricardo D.~Katz for helpful discussions on the topic. The authors are grateful to the anonymous reviewers of CPP'23 for their comments and suggestions. 
\end{acks}

\bibliographystyle{ACM-Reference-Format}

\end{document}